\documentclass[english,notitlepage,reprint,aip]{revtex4-2}
\usepackage[T1]{fontenc}
\usepackage[latin9]{inputenc}
\usepackage{mathrsfs}
\usepackage{amsmath}
\usepackage{amsthm}
\usepackage{amssymb}
\usepackage{graphicx}

\makeatletter

\providecommand{\tabularnewline}{\\}

\makeatother

\usepackage{babel}
\begin{document}
\title{Weak second-order quantum state diffusion unraveling of the Lindblad
master equation }
\author{Sayak Adhikari}
\address{Fritz Haber Center for Molecular Dynamics and Institute of Chemistry,
The Hebrew University of Jerusalem, Jerusalem 9190401, Israel}
\author{Roi Baer}
\email{roi.baer@mail.huji.ac.il}

\address{Fritz Haber Center for Molecular Dynamics and Institute of Chemistry,
The Hebrew University of Jerusalem, Jerusalem 9190401, Israel}
\begin{abstract}
Simulating mixed-state evolution in open quantum systems is crucial
for various chemical physics, quantum optics, and computer science
applications. These simulations typically follow the Lindblad master
equation dynamics. An alternative approach known as quantum state
diffusion unraveling is based on the trajectories of pure states generated
by random wave functions, which evolve according to a nonlinear It\^{o}-Schr\"{o}dinger
equation (ISE). This study introduces weak first- and second-order
solvers for the ISE based on directly applying the It\^{o}-Taylor expansion
with exact derivatives in the interaction picture. We tested the method
on free and driven Morse oscillators coupled to a thermal environment
and found that both orders allowed practical estimation with a few
dozen iterations. The variance was relatively small compared to the
linear unraveling and did not grow with time. The second-order solver
delivers much higher accuracy and stability with bigger time steps
than the first-order scheme, with a small additional workload. However,
the second-order algorithm has quadratic complexity with the number
of Lindblad operators as opposed to the linear complexity of the first-order
algorithm.
\end{abstract}
\maketitle

\section{\label{sec:Introduction}Introduction}

When a physical system in a pure quantum state is brought to interact
weakly with a macroscopic thermal environment, it changes its energy
and chemical composition. At the same time, it gradually loses its
\textquotedbl quantumness\textquotedbl{} or, more technically, its
phase coherence. Ultimately, the system's state resembles that drawn
randomly from the Gibbs ensemble at the environment's temperature
and chemical potentials. All quantum systems interact with the environment.
Therefore, techniques to simulate decoherence and decay processes
are vital for developing quantum technologies and studying chemical
processes in solutions and condensed matter. \citep{pollard1994solution,kohen1997phasespace,nitzan2006chemical,harbola2006quantum,breuer2007thetheory,appel2009stochastic,rivas2012openquantum,biele2012astochastic,blum2012density,schaller2014openquantum,gardiner2014thequantum,uzdin2016speedlimits,alicki2018introduction,ruan2018unravelling,kurizki2021thermodynamics,levy2021response,gerry2023fullcounting}

The pure quantum state of an open system is not known with certainty,
and thus, we consider it a random mixture of pure states. The density
operator $\rho$ is the mathematical object that best describes this
mixture, enabling the calculation of probabilities of outcomes of
measurements. Even when the initial mixture $\rho\left(0\right)$
is known, the density operator $\rho\left(t\right)$ changes over
time. The Redfield master equation \citep{redfield1957onthe,pollard1994solution,gaspard1999slippage,nitzan2006chemical,esposito2010selfconsistent,becker2021lindbladian}
is one way to approximate this, but it sometimes creates mixtures
with negative probabilities. Lindblad's master equation \citep{lindblad1976onthe,gorini1976completely,alicki2007quantum,alicki2018introduction,manzano2020ashort}
is an augmented form of Redfield's equation, guaranteeing the density
operator's positivity. It is a quantum Liouville-like equation but
includes additional terms, relying on \emph{Lindblad operators}, to
represent the dressed system-environment interactions.

The density operator of the Lindblad equation can be modeled by stochastic
processes collectively called \textquotedbl quantum unraveling models.\textquotedbl{}
\citep{plenio1998thequantumjump,percival1998quantum,breuer2007thetheory}
They provide recipes for generating a random time-dependent normalized
pure state $\left|\psi\left(t\right)\right\rangle $ for which the
expected value of the projector, $\mathbb{E}\left[\left|\psi\left(t\right)\right\rangle \left\langle \psi\left(t\right)\right|\right]$,
is identically equal to the Lindblad density operator $\rho\left(t\right)$.
One type of unraveling is the Monte-Carlo wave function approach \citep{dalibard1992wavefunction,gardiner1992wavefunction,carmichael1993anopen},
also known as the \textquotedbl quantum jumps model,\textquotedbl{}
where the Lindblad operators operate as \textquotedbl jump operators.\textquotedbl{}
A second approach to unraveling is the \textquotedbl quantum state
diffusion model\textquotedbl{} \citep{gisin1992thequantumstate},
involving a norm-conserving (but not unitary) time-dependent stochastic
It\^{o}-Schr\"{o}dinger equation (ISE) for $\left|\psi\left(t\right)\right\rangle $.
The ISE contains drift (evolution) and diffusion (fluctuation) terms.
The quantum jump and quantum state diffusion models yield different
trajectories: the former evolves non-continuously. At the same time,
the latter is continuous but non-differentiable in time. 

One advantage of basing numerical simulations on the quantum state
diffusion model is the availability of well-established high-order
techniques for solving stochastic differential equations (SDEs) \citep{breuer2007thetheory,li2020exponential,mora2023onthe,johansson2013qutip2}.
In the present contribution, we deploy a simple approach based on
exact derivatives in the interaction picture, an It\^{o}-Taylor expansion
for weak second-order solutions. The method is stable and allows for
high accuracy and slight variance. 

\section{Weak second-order quantum state diffusion unraveling }

\subsection{Comments on notation}

Before we start the detailed theory, here are several comments concerning
the notation in this paper: 
\begin{enumerate}
\item The time dimension of any quantity can be read-off from its superscripts
or subscripts: a subscript $0$ adds a dimension of $time^{-1}$ and
a superscript $0$ attributes a dimension $time^{+1}$. Thus, the
Hamiltonian $\mathcal{H}_{0}$ has the dimension of inverse time while
the symbol $I^{0}$ has the dimension of time. A Greek subscript attributes
an additional factor of $time^{-1/2}$ and a Greek superscript an
additional factor of $time^{1/2}$. Thus, the symbol $I^{\alpha}$
has the dimension of $times^{1/2}$ while $I_{\alpha\beta}$ has the
dimension of $time^{-1}.$ The Kronecker-delta $\delta_{\alpha}^{\beta}$
is dimensionless. Furthermore, the symbols $I_{\alpha}^{0}$ and $I_{\gamma}^{\alpha\beta}$
have the dimension of $time^{1/2}$ while $I^{0\alpha}$ has the dimensions
of $time^{3/2}$. This convention helps to ascertain that the different
time orders we use in our analytical developments are consistent (i.e.
that we do not add quantities with different time dimensions). 
\item The index $\alpha$, $\alpha'$, going from $1,\dots,N_{L}$ denotes
one of the $N_{L}$ Lindblad operators. When two quantities indexed
with $\alpha$ are multiplied in an expression, a summation over $\alpha$
from 1 to $N_{L}$ is assumed and we omit the explicit $\sum_{\alpha=1}^{N_{L}}$
notation (this is the so-called Einstein convention). If the index
is decorated by a dot $\dot{\alpha}$ then no such summation is implied.
\item Below we introduce a ``0'' operator, in addition to the Lindblad
operators. Unlike the $\alpha$, $\alpha'$ indices discussed above,
going from $1,\dots,N_{L}$, we also use the $\beta$, $\beta'$ indices
to enumerate operators and quantities that range from $0$ to $N_{L}$.
Similar to the case with $\alpha$ , when two quantities indexed with
$\beta$ are multiplied in an expression, a summation over $\beta$
is assumed and we omit the explicit $\sum_{\beta=0}^{N_{L}}$ notation.
If the index is decorated by a dot $\dot{\beta}$ then no such summation
is implied.
\end{enumerate}

\subsection{Quantum state diffusion unraveling}

The Lindblad equation
\begin{equation}
\dot{\rho}\left(t\right)=-i\left[\mathcal{H}_{0}+\theta\left(t\right)\mathcal{V}_{0},\rho\right]+\mathscr{D}_{0}\rho\label{eq:Lindbald}
\end{equation}
together with the initial condition $\rho\left(0\right)$, determines
$\rho\left(t\right)$ for all time $t>0$. It contains unitary terms
dependent on $\mathcal{H}_{0}$, an effective Hamiltonian operator,
and $\theta\left(t\right)\mathcal{V}_{0}$ a driving force with $\theta\left(t\right)$
a dimensionless real time-dependent envelop with time derivative $\theta_{0}\left(t\right)\equiv\dot{\theta}\left(t\right)$.
It also contains dissipative terms \citep{gisin1992thequantumstate,alicki2007quantum,manzano2020ashort,alicki2018introduction}:

\begin{align}
\mathscr{D}_{0}\rho & \equiv\left[\mathcal{L}_{\alpha}\rho,\mathcal{L}_{\alpha}^{\dagger}\right]+\left[\mathcal{L}_{\alpha},\rho\mathcal{L}_{\alpha}^{\dagger}\right],\label{eq:Dissipator}
\end{align}
defined in terms the Lindblad operators $\mathcal{L}_{\alpha}$, $\alpha=1,\dots,N_{L}$.
Atomic units are used ($\hbar=1$, $m_{e}=1$) here, so the energy
and inverse time units are identical. Accordingly, $\mathcal{L}_{\alpha}$
have the dimension of $time^{-1/2}$.  

Evolving the mixed state density operator $\rho\left(t\right)$ using
Eq.~(\ref{eq:Lindbald}) can be numerically expensive when systems
are large. A possible simplification can be achieved by the unraveling
procedure, which evolves a pure random state $\left|\psi\left(t\right)\right\rangle $
in such a way that $\mathbb{E}\left[\left|\psi\left(t\right)\right\rangle \left\langle \psi\left(t\right)\right|\right]=\rho\left(t\right)$.
 In quantum state diffusion unraveling $\left|\psi\left(t\right)\right\rangle $
is obtained from the following It\^{o}-Schr\"{o}dinger equation (ISE) \citep{gisin1992thequantumstate}
\begin{align}
\left|d\psi\right\rangle  & =-i\mathcal{H}_{0}\left|\psi\right\rangle dw^{0}+\Lambda_{\beta}\left|\psi\right\rangle dw^{\beta},\label{eq:SSE}
\end{align}
starting from a random ket $\left|\psi\left(0\right)\right\rangle $
for which $\mathbb{E}\left[\left|\psi\left(0\right)\right\rangle \left\langle \psi\left(0\right)\right|\right]=\rho\left(0\right)$.
In Eq.~(\ref{eq:SSE}),
\begin{align*}
\Lambda_{\alpha} & \equiv\mathcal{L}_{\alpha}-\left\langle \mathcal{L}_{\alpha}\right\rangle \\
\Lambda_{0} & \equiv-i\mathcal{\theta}\left(t\right)\mathcal{V}_{0}\left(t\right)+\left(2\left\langle \mathcal{L}_{\alpha}^{\dagger}\right\rangle \mathcal{L}_{\alpha}-\mathcal{L}_{\alpha}^{\dagger}\mathcal{L}_{\alpha}-\left\langle \mathcal{L}_{\alpha}^{\dagger}\right\rangle \left\langle \mathcal{L}_{\alpha}\right\rangle \right),
\end{align*}
and
\begin{equation}
\left\langle \mathcal{L}_{\alpha}\right\rangle \equiv\frac{\left\langle \psi\left|\mathcal{L}_{\alpha}\right|\psi\right\rangle }{\left\langle \psi\left|\psi\right.\right>}.\label{eq:ExpectationValue}
\end{equation}
Notice that $\left\langle \Lambda_{\alpha}\right\rangle =0$ (for
$\alpha=1,\dots,N_{L}$). In the above, $dw^{0}=dt$ is the time-step
while $dw^{\alpha}$, $\alpha=1,2,\dots N_{L}$ are independent complex
Wiener processes, with real $\Re\left[dw^{\alpha}\right]$ and imaginary
$\Im\left[dw^{\alpha}\right]$ parts, each of which is an independent
\emph{real} Wienner process with zero expected value and a variance
equal to $dt$. As is common in the stochastic differential equations
literature we omit the expected value symbol $\mathbb{E}$ from differentials
hence we are lead to the following variances for $dw^{\alpha}$:
\begin{equation}
\left(dw^{\alpha}\right)^{2}=\left(dw^{\alpha*}\right)^{2}=0,\quad\left|dw^{\alpha}\right|^{2}=2dt.\label{eq:dw-relations}
\end{equation}
Note that $dw^{\alpha}$ are also independent of $\left|\psi\right\rangle $.
Note, that the differential $d\left\langle \psi\left|\psi\right.\right>\equiv\left\langle d\psi\left|\psi\right.\right>+\left\langle \psi\left|d\psi\right.\right>+\left\langle d\psi\left|d\psi\right.\right>$
vanishes when evaluated using Eqs.~(\ref{eq:SSE})-(\ref{eq:dw-relations}).
Hence $\left\langle \psi\left|\psi\right.\right>$ is a constant of
motion, separately for each trajectory. 

\subsection{Weak first- and second-order propagators}

The first step in providing a solution to the ISE, is to move to the
interaction picture, defining $\left|\phi\left(t\right)\right\rangle \equiv e^{i\mathcal{H}_{0}t}\left|\psi\left(t\right)\right\rangle $
and for any operator $\mathcal{Y}$, $\mathcal{Y}\left(t\right)\equiv e^{i\mathcal{H}_{0}t}\mathcal{Y}e^{-i\mathcal{H}_{0}t}$.
The ISE of Eq.~(\ref{eq:SSE}) becomes
\begin{equation}
d\left|\phi\right\rangle =dw^{\beta}\Lambda_{\beta}\left(t\right)\left|\phi\right\rangle \label{eq:phi-sse}
\end{equation}
where 
\begin{equation}
\Lambda_{\alpha}\left(t\right)\equiv\mathcal{L}_{\alpha}\left(t\right)-\left\langle \mathcal{L}_{\alpha}\left(t\right)\right\rangle ,
\end{equation}
and note our definition of the expectation value
\[
\left\langle \mathcal{L}_{\alpha}\left(t\right)\right\rangle \equiv\frac{\left\langle \psi\left(t\right)\left|\mathcal{L}_{\alpha}\right|\psi\left(t\right)\right\rangle }{\left\langle \psi\left(t\right)\left|\psi\left(t\right)\right.\right>}=\frac{\left\langle \phi\left(t\right)\left|\mathcal{L}_{\alpha}\left(t\right)\right|\phi\left(t\right)\right\rangle }{\left\langle \phi\left(t\right)\left|\phi\left(t\right)\right.\right>},
\]
which includes division by the norm and thus different from some other
applications (e.g., \citep{mora2023onthe}). Formally there is no
need to divide by the norm, since one can choose the initial norm
as 1 and it is preserved. However, in practice the norm is never perfectly
preserved so this division is not a trivial change and we found that
division by the norm leads to a more stable numerical behavior. 

For developing the numerical scheme, we divide time $t\in\left[0,T_{f}\right]$
where $T_{f}$ is the final time, into $N_{T}$ discrete small temporal
segments $\Delta T=T_{f}/N_{T}$, and designate $t_{n+1}=t_{0}+n\Delta T$,
$n=1,2,\dots,N_{T}$. Using the notation $\left|\Phi\right\rangle \equiv\left|\phi\left(t_{n}\right)\right\rangle $,
the change in the evolving ket during the $n$th time step, $\left|\Delta\Phi\right\rangle \equiv\left|\phi\left(t_{n+1}\right)\right\rangle -\left|\Phi\right\rangle $,
is expressed as a stochastic integral over $d\left|\phi\right\rangle $,
which gives, using Eq.~(\ref{eq:phi-sse}):
\begin{align}
\left|\Delta\Phi\right\rangle  & =\int_{t_{n}}^{t_{n}+\Delta T}\Lambda_{\beta}\left(\tau\right)\left|\phi\left(\tau\right)\right\rangle dw_{\tau}^{\beta}.\label{eq:DeltaPhi}
\end{align}
We strive for an approximation of this integral, which allows an exact
solution of the ISE in the limit of $N_{T}\to\infty$ and, accordingly,
$\Delta T\to0$. Our analysis follows closely that found in the classical
literature on numerical solutions of real SDEs \citep{kloeden1992numerical,milstein1995numerical}.
Our contribution is the adaptation of the theory to Eq.~\ref{eq:phi-sse},
including the use of complex Wiener processes and exact analytical
derivatives therein. We also contribute a simplified notation scheme. 

The change in the wave function is provided in terms of first- and
second-order contributions, $\left|\Delta\Phi\right\rangle \approx\left|\Delta^{\left(1\right)}\Phi\right\rangle +\left|\Delta^{\left(2\right)}\Phi\right\rangle $.
The first-order term is obtained by approximating $\Lambda_{\beta}\left(\tau\right)\left|\phi\left(\tau\right)\right\rangle $
as $\left|\beta\right\rangle \equiv\Lambda_{\beta}\left(t_{n}\right)\left|\Phi\right\rangle $
for $\tau\in\left[t_{n},t_{n}+\Delta T\right]$. This gives: 
\begin{equation}
\left|\Delta^{\left(1\right)}\Phi\right\rangle =I^{\beta}\left|\beta\right\rangle ,\label{eq:First-Order-Correction}
\end{equation}
where $I^{\alpha}=\int_{t_{n}}^{t_{n}+\Delta T}dw^{\alpha}$, $\alpha=1,\dots,N_{L}$
are It� integrals given in Table~(\ref{tab:Stoch-Integs}) and $I^{0}=\Delta T$.
In the numerical calculations we use the model for the complex stochastic
It� integrals given in the last column of the table.  

We use the It�-Taylor expansion to the lowest order for the second-order
correction. For this, we introduce a notation in which all quantities
are first written as functions of a ket $\left|x\right\rangle $ and
a (different) bra $\left\langle y\right|$, then we take separate
derivatives with respect to them, and only after that do we set $\left|x\right\rangle =\left|\Phi\right\rangle $
and $\left\langle y\right|=\left\langle \Phi\right|$. In the supplementary
information we give a detailed explanation of the results we present
here. We define, for $\alpha=1,\dots,N_{L}$ the $\ell$-functions
of $\left|x\right\rangle $, $\left\langle y\right|$ and the time
$t$, 
\[
\ell_{\alpha}\left(\left|x\right\rangle ,\left\langle y\right|,t\right)\equiv\frac{\left\langle y\left|\mathcal{L}_{\alpha}\left(t\right)\right|x\right\rangle }{\left\langle y\left|x\right.\right>},
\]
and
\[
\ell_{\alpha}^{*}\left(\left|x\right\rangle ,\left\langle y\right|,t\right)\equiv\frac{\left\langle y\left|\mathcal{L}_{\alpha}^{\dagger}\left(t\right)\right|x\right\rangle }{\left\langle y\left|x\right.\right>},
\]
which, when evaluated at $\Phi$, become the expectation values of
the Lindblad operators:
\begin{align*}
\left(\ell_{\alpha}\left(\left|x\right\rangle ,\left\langle y\right|,t\right)\right)_{\Phi} & \equiv\ell_{\alpha}\left(\left|\Phi\right\rangle ,\left\langle \Phi\right|,t\right)=\left\langle \mathcal{L}_{\alpha}\left(t\right)\right\rangle ,\\
\left(\ell_{\alpha}^{*}\left(\left|x\right\rangle ,\left\langle y\right|,t\right)\right)_{\Phi} & \equiv\ell_{\alpha}^{*}\left(\left|\Phi\right\rangle ,\left\langle \Phi\right|,t\right)=\left\langle \mathcal{L}_{\alpha}^{\dagger}\left(t\right)\right\rangle .
\end{align*}
The derivative of $\ell_{\alpha}\left(\left|x\right\rangle ,\left\langle y\right|,t\right)$
with respect to the bra $\left\langle y\right|$results in a ket:
\begin{align}
\left|\frac{\partial}{\partial\left\langle y\right|}\ell_{\alpha}\left(\left|x\right\rangle ,\left\langle y\right|,t\right)\right\rangle  & =\frac{\left(\mathcal{L}_{\alpha}\left(t\right)-\ell_{\alpha}\left(\left|x\right\rangle ,\left\langle y\right|,t\right)\right)\left|x\right\rangle }{\left\langle y\left|x\right.\right>}\nonumber \\
 & \equiv\frac{\left|\lambda_{\alpha}\left(\left|x\right\rangle ,\left\langle y\right|,t\right)\right\rangle }{\left\langle y\left|x\right.\right>},\label{eq:lambda-alpha}
\end{align}
which is orthogonal to $\left|y\right\rangle $: 
\[
\left\langle y\left|\lambda_{\alpha}\right.\right>=0.
\]
Similarly, the derivative with respect to the ket $\left|x\right\rangle $
results in the bra:
\begin{align*}
\left\langle \frac{\partial}{\partial\left|x\right\rangle }\ell_{\alpha}\left(\left|x\right\rangle ,\left\langle y\right|,t\right)\right| & =\frac{\left\langle y\right|\left(\mathcal{L}_{\alpha}\left(t\right)-\ell_{\alpha}\left(\left|x\right\rangle ,\left\langle y\right|,t\right)\right)}{\left\langle y\left|x\right.\right>}\\
 & \equiv\left\langle \mu_{\alpha}\left(\left|x\right\rangle ,\left\langle y\right|,t\right)\right|,
\end{align*}
which is orthogonal to $\left|x\right\rangle $$:$
\[
\left\langle \mu_{\alpha}\left|x\right.\right>=0.
\]

We extend the definition of the '$\lambda$-kets', by adding a ``zero''
subscript:\onecolumngrid
\begin{align}
\left|\lambda_{0}\left(\left|x\right\rangle ,\left\langle y\right|,t\right)\right\rangle  & \equiv-i\mathcal{V}_{0}\left(t\right)\mathcal{\theta}\left(t\right)\left|x\right\rangle +\left(2\ell_{\alpha}^{*}\left(\left|x\right\rangle ,\left\langle y\right|,t\right)\mathcal{L}_{\alpha}-\mathcal{L}_{\alpha}^{\dagger}\mathcal{L}_{\alpha}-\ell_{\alpha}\left(\left|x\right\rangle ,\left\langle y\right|,t\right)\ell_{\alpha}^{*}\left(\left|x\right\rangle ,\left\langle y\right|,t\right)\right)\left|x\right\rangle .\label{eq:lambda0}
\end{align}
\twocolumngrid

When evaluated at $\Phi$ we have, for $\beta=0,\dots,N_{L}$:
\[
\left|\lambda_{\beta}\left(\left|x\right\rangle ,\left\langle y\right|,t\right)\right\rangle _{\Phi,t_{n}}\equiv\Lambda_{\beta}\left|\Phi\right\rangle \equiv\left|\beta\right\rangle .
\]
With these definitions, the second-order correction is given in terms
of the $\lambda$-kets $t$, $\left|x\right\rangle $, $\left\langle y\right|$
first derivative and the $\left|x\right\rangle \left\langle y\right|$
mixed derivatives as follows:   \onecolumngrid

\begin{align}
\left|\Delta^{\left(2\right)}\Phi\right\rangle  & =I^{0\beta}\left(\frac{\partial}{\partial t}\left|\lambda_{\beta}\right\rangle \right)_{\Phi,t_{n}}+\underset{'X'}{\underbrace{I^{\beta\beta'}\left(\frac{\partial}{\partial\left|x\right\rangle }\left|\lambda_{\beta'}\right\rangle \right)_{\Phi,t_{n}}\left|\beta\right\rangle }}\label{eq:Second-Order-Correction}\\
 & +\underset{'Y'}{\underbrace{I^{\beta*\beta'}\left\langle \beta\right|\left(\frac{\partial}{\partial\left\langle y\right|}\left|\lambda_{\beta'}\right\rangle \right)_{\Phi,t_{n}}}}+\underset{'XY'}{\underbrace{2I^{0\beta}\left\langle \alpha\right|\left(\frac{\partial^{2}}{\partial\left|x\right\rangle \partial\left\langle y\right|}\left|\lambda_{\beta}\right\rangle \right)_{\Phi,t_{n}}\left|\alpha\right\rangle }}\nonumber 
\end{align}
where $I^{\beta\beta'}$ ($\beta=0,\dots,N_{L}$, $\alpha'=0,\dots,N_{L}$)
are the It� integrals defined in Table \ref{tab:Stoch-Integs}. 

\begin{table*}
\caption{\label{tab:Stoch-Integs}The definition of the stochastic It� integrals
used in Eqs.~(\ref{eq:First-Order-Correction})-(\ref{eq:Second-Order-Correction}),
where $t_{n}$ are the propagation time steps, with $t_{n+1}-t_{n}=\Delta T$,
and $\alpha,\alpha',\alpha'',\alpha'''=1,\dots,N_{L}$ are Lindblad
indices and $\alpha*\equiv\alpha+N_{L}$, etc. All the integrals have
zero expected value and covariance described in the table. The integrals
$I^{0}$ and $I^{00}$ are deterministic and equal to $\Delta T$
and $\frac{\Delta T^{2}}{2}$, respectively. The last column for each
integral gives a model depending on $4\times N_{L}$ independent complex
random variables $m^{\alpha}$ ($\alpha=1,\dots,N_{L}$, $m=a,b,c,d$),
distributed with $\mathbb{E}\left[m^{\alpha}\right]=0$, $\mathbb{E}\left[m^{\alpha}m'^{\alpha'}\right]=0$,
and $\mathbb{E}\left[m^{\alpha*}m'^{\alpha'}\right]=2\Delta T\delta_{\alpha\alpha'}\delta_{mm'}$.
 For each time interval $t_{n}\to t_{n+1}$ a new uncorrelated set
of such random variables is used.}
\begin{tabular}{cccccc||c}
\hline 
Integral & $I^{\alpha'*}$ & $I^{\alpha'*0}$ & $I^{0\alpha'*}$ & $I^{\alpha''*\alpha'''*}$ & $I^{\alpha''\alpha'''*}$ & Model\tabularnewline
\hline 
\hline 
\noalign{\vskip\doublerulesep}
$I^{\alpha}\equiv\int_{0}^{\Delta T}dw_{\tau}^{\alpha}$ & $\delta_{\alpha'}^{\alpha}2\Delta T$ & $\delta_{\alpha'}^{\alpha}\Delta T^{2}$ & $\delta_{\alpha'}^{\alpha}\Delta T^{2}$ & 0 & 0 & $a^{\alpha}$\tabularnewline[\doublerulesep]
\hline 
\noalign{\vskip\doublerulesep}
$I^{\alpha0}\equiv\int_{0}^{\Delta T}\int_{0}^{\tau}dw_{\tau'}^{\alpha}d\tau$ & $\delta_{\alpha'}^{\alpha}\Delta T^{2}$ & $\delta_{\alpha'}^{\alpha}\frac{2\Delta T^{3}}{3}$ & $\delta_{\alpha'}^{\alpha}\frac{\Delta T^{3}}{3}$ & 0 & 0 & $\left(a^{\alpha}+\frac{1}{\sqrt{3}}b^{\alpha}\right)\frac{\Delta T}{2}$\tabularnewline[\doublerulesep]
\hline 
\noalign{\vskip\doublerulesep}
$I^{0\alpha}\equiv\int_{0}^{\Delta T}\left(\tau-t_{n}\right)dw_{\tau}^{\alpha}$ & $\delta_{\alpha'}^{\alpha}\Delta T^{2}$ & $\delta_{\alpha'}^{\alpha}\frac{\Delta T^{3}}{3}$ & \textbf{$\delta_{\alpha'}^{\alpha}\frac{2\Delta T^{3}}{3}$} & 0 & 0 & $\left(a^{\alpha}-\frac{1}{\sqrt{3}}b^{\alpha}\right)\frac{\Delta T}{2}$\tabularnewline[\doublerulesep]
\hline 
\noalign{\vskip\doublerulesep}
$I^{\alpha\alpha'}\equiv\int_{0}^{\Delta T}dw_{\tau}^{\alpha}\int_{0}^{\tau}dw_{\tau'}^{\alpha'}$ & 0 & 0 & 0 & $\delta_{\alpha''}^{\alpha}\delta_{\alpha'}^{\alpha'''}2\Delta T^{2}$ & 0 & $\frac{1}{\sqrt{2}}c^{\alpha}d^{\alpha'}$\tabularnewline[\doublerulesep]
\hline 
\noalign{\vskip\doublerulesep}
$I^{\alpha*\alpha'}\equiv\int_{0}^{\Delta T}dw_{\tau}^{\alpha*}\int_{0}^{\tau}dw_{\tau'}^{\alpha'}$ & 0 & 0 & 0 & 0 & $\delta_{\alpha''}^{\alpha}\delta_{\alpha'}^{\alpha'''}2\Delta T^{2}$ & $\frac{1}{\sqrt{2}}\left(c^{\alpha}\right)^{*}d^{\alpha'}$\tabularnewline[\doublerulesep]
\hline 
\end{tabular}
\end{table*}
 The derivative in the expression for $\left|\Delta\Phi^{\left(2\right)}\right\rangle $
are: 
\begin{align*}
\left(\frac{\partial}{\partial t}\left|\lambda_{\alpha}\right\rangle \right)_{\Phi,t_{n}} & =i\left(\left[\mathcal{H}_{0},\Lambda_{\alpha}\right]-\left\langle \left[\mathcal{H}_{0},\Lambda_{\alpha}\right]\right\rangle \right)\left|\Phi\right\rangle ,\\
\left(\frac{\partial}{\partial t}\left|\lambda_{0}\right\rangle \right)_{\Phi,t_{n}}= & \left(\theta\left(t_{n}\right)\left[\mathcal{H}_{0},\mathcal{V}_{0}\right]-i\theta_{0}\left(t_{n}\right)\mathcal{V}_{0}\right)\left|\Phi\right\rangle +i\left(2\left\langle \left[\mathcal{H}_{0},\Lambda_{\alpha}^{\dagger}\right]\right\rangle \Lambda_{\alpha}-\left[\mathcal{H}_{0},\Lambda_{\alpha}^{\dagger}\Lambda_{\alpha}\right]\right)\left|\Phi\right\rangle \\
 & +i\left\langle \mathcal{L}_{\alpha}^{\dagger}\right\rangle \left(\left[\mathcal{H}_{0},\Lambda_{\alpha}\right]-\left\langle \left[\mathcal{H}_{0},\Lambda_{\alpha}\right]\right\rangle \right)\left|\Phi\right\rangle .
\end{align*}
Next, using the notation $\left|\beta\right\rangle \equiv\Lambda_{\beta}\left|\Phi\right\rangle $,
$\left|\alpha\beta\right\rangle \equiv\Lambda_{\alpha}\Lambda_{\beta}\left|\Phi\right\rangle $
etc., the $x$-derivatives are:

\begin{align*}
\left(\frac{\partial}{\partial\left|x\right\rangle }\left|\lambda_{\alpha}\right\rangle \right)_{\Phi_{n}}\left|\beta\right\rangle  & =\left|\alpha\beta\right\rangle -\left|\Phi\right\rangle \left\langle \Phi\left|\alpha\beta\right.\right>,\\
\left(\frac{\partial}{\partial\left|x\right\rangle }\left|\lambda_{0}\right\rangle \right)_{\Phi_{n}}\left|\beta\right\rangle  & =\left|0\beta\right\rangle +\sum_{\alpha=1}^{N_{L}}\left(\left(2\left|\alpha\right\rangle +\left|\Phi\right\rangle \left\langle \mathcal{L}_{\alpha}\right\rangle \right)\left\langle \alpha\left|\beta\right.\right>-\left|\Phi\right\rangle \left\langle \Phi\left|\alpha\beta\right.\right>\left\langle \mathcal{L}_{\alpha}^{\dagger}\right\rangle \right),
\end{align*}
the y-derivatives are:
\begin{align*}
\left\langle \beta\right|\left(\frac{\partial}{\partial\left\langle y\right|}\left|\lambda_{\alpha}\right\rangle \right)_{\Phi_{n}} & =-\left|\Phi\right\rangle \left\langle \beta\left|\alpha\right.\right>,\\
\left\langle \beta\right|\left(\frac{\partial}{\partial\left\langle y\right|}\left|\lambda_{0}\right\rangle \right)_{\Phi_{n}} & =\left(\left(\left|\alpha\right\rangle 2+\left|\Phi\right\rangle \left\langle \mathcal{L}_{\alpha}\right\rangle \right)\left\langle \beta\alpha\left|\Phi\right.\right>-\left|\Phi\right\rangle \left\langle \beta\left|\alpha\right.\right>\left\langle \mathcal{L}_{\alpha}^{\dagger}\right\rangle \right),
\end{align*}
 and the mixed derivatives are: 
\begin{align*}
\left\langle \alpha\right|\left(\frac{\partial^{2}}{\partial\left|x\right\rangle \partial\left\langle y\right|}\left|\lambda_{\alpha'}\right\rangle \right)_{\Phi_{n}}\left|\alpha\right\rangle  & =-\left(\left|\alpha\right\rangle \left\langle \alpha\left|\alpha'\right.\right>+\left|\Phi\right\rangle \left\langle \alpha\left|\alpha'\alpha\right.\right>\right),
\end{align*}

\begin{align*}
\left\langle \alpha\right|\left(\frac{\partial^{2}}{\partial\left|x\right\rangle \partial\left\langle y\right|}\left|\lambda_{0}\right\rangle \right)_{\Phi_{n}}\left|\alpha\right\rangle  & =\left|\alpha\right\rangle \left(2\left\langle \alpha'\alpha\left|\alpha'\right.\right>+\left\langle \mathcal{L}_{\alpha'}\right\rangle \left\langle \alpha'\alpha\left|\Phi\right.\right>-\left\langle \alpha'\left|\alpha\right.\right>\left\langle \mathcal{L}_{\alpha'}^{\dagger}\right\rangle \right)+2\left|\alpha'\alpha\right\rangle \left\langle \alpha\alpha'\left|\Phi\right.\right>\\
 & \qquad-\left|\Phi\right\rangle \left(\left|\left\langle \alpha\left|\alpha'\right.\right>\right|^{2}+\left|\left\langle 0\left|\alpha'\alpha\right.\right>\right|^{2}+2i\Im\left[\left\langle \alpha\alpha'\left|\alpha\right.\right>\left\langle \mathcal{L}_{\alpha'}\right\rangle \right]\right).
\end{align*}
A further simplification is obtained using the following summed kets:
\begin{align*}
\left|e^{0}\right\rangle  & \equiv I^{0\alpha}\left|\alpha\right\rangle ,\qquad\left|f^{0}\right\rangle \equiv I^{\alpha0}\left|\alpha\right\rangle ,\qquad\left|f^{0*}\right\rangle \equiv I^{\alpha0*}\left|\alpha\right\rangle \\
\left|c\right\rangle  & \equiv c^{\alpha}\left|\alpha\right\rangle ,\qquad\left|d\right\rangle \equiv d^{\alpha}\left|\alpha\right\rangle ,\qquad\left|d^{*}\right\rangle \equiv d^{\alpha*}\left|\alpha\right\rangle ,\qquad\left|dc\right\rangle \equiv d^{\alpha'}\Lambda_{\alpha'}\left|c\right\rangle ,
\end{align*}
with which the $'X'$, $'Y'$ and $'XY'$ terms of Eq.~(\ref{eq:Second-Order-Correction})
become:
\begin{align*}
'X' & =\frac{\Delta T^{2}}{2}\left(\left|00\right\rangle +\left(\left(2\left|\alpha\right\rangle +\left|\Phi\right\rangle \left\langle \mathcal{L}_{\alpha}\right\rangle \right)\left\langle \alpha\left|0\right.\right>-\left|\Phi\right\rangle \left\langle \Phi\left|\alpha0\right.\right>\left\langle \mathcal{L}_{\alpha}^{\dagger}\right\rangle \right)\right)+\left|0f^{0}\right\rangle +\left|e^{0}0\right\rangle \\
 & +\left(\left(2\left|\alpha\right\rangle +\left|\Phi\right\rangle \left\langle \mathcal{L}_{\alpha}\right\rangle \right)\left\langle \alpha\left|f^{0}\right.\right>-\left|\Phi\right\rangle \left\langle \Phi\left|\alpha f^{0}\right.\right>\left\langle \mathcal{L}_{\alpha}^{\dagger}\right\rangle \right)-\left|\Phi\right\rangle \left\langle \Phi\left|e^{0}0\right.\right>+\left|dc\right\rangle -\left|\Phi\right\rangle \left\langle \Phi\left|dc\right.\right>,
\end{align*}

\begin{align*}
'Y' & =\frac{\Delta T^{2}}{2}\left(\left(\left|\alpha\right\rangle 2+\left|\Phi\right\rangle \left\langle \mathcal{L}_{\alpha}\right\rangle \right)\left\langle 0\alpha\left|\Phi\right.\right>-\left|\Phi\right\rangle \left\langle 0\left|\alpha\right.\right>\left\langle \mathcal{L}_{\alpha}^{\dagger}\right\rangle \right)\\
 & +\left[\left(\left|\alpha'\right\rangle 2+\left|\Phi\right\rangle \left\langle \mathcal{L}_{\alpha'}\right\rangle \right)\left\langle f^{0*}\alpha'\left|\Phi\right.\right>-\left|\Phi\right\rangle \left\langle f^{0*}\left|\alpha'\right.\right>\left\langle \mathcal{L}_{\alpha'}^{\dagger}\right\rangle \right]-\left|\Phi\right\rangle \left\langle 0\left|e^{0}\right.\right>-\left|\Phi\right\rangle \left\langle d^{*}\left|c\right.\right>,
\end{align*}
\begin{align}
'XY' & =-2\left(\left|\alpha\right\rangle \left\langle \alpha\left|e^{0}\right.\right>+\left|\Phi\right\rangle \left\langle \alpha\left|e^{0}\alpha\right.\right>\right)\nonumber \\
 & +\left|\alpha\right\rangle \left(2\left\langle \alpha'\alpha\left|\alpha'\right.\right>+\left\langle \mathcal{L}_{\alpha'}\right\rangle \left\langle \alpha'\alpha\left|\Phi\right.\right>-\left\langle \alpha'\left|\alpha\right.\right>\left\langle \mathcal{L}_{\alpha'}^{\dagger}\right\rangle \right)\Delta T^{2}\nonumber \\
 & +\left|\alpha'\alpha\right\rangle \left(2\left\langle \alpha\alpha'\left|\Phi\right.\right>\right)\Delta T^{2}\nonumber \\
 & -\left|\Phi\right\rangle \left(\left|\left\langle \alpha\left|\alpha'\right.\right>\right|^{2}+\left|\left\langle 0\left|\alpha'\alpha\right.\right>\right|^{2}+2i\Im\left[\left\langle \alpha\alpha'\left|\alpha\right.\right>\left\langle \mathcal{L}_{\alpha'}\right\rangle \right]\right)\Delta T^{2}.\label{eq:XY-expression}
\end{align}

\twocolumngrid

This completes the description of the method. As for the algorithmic
scaling, the evaluation of each of the terms in the 'X' and 'Y' expressions
requires order-$N_{L}$ operations (linear scaling effort in the number
of Lindblad operators). However, the 'XY' expression includes terms
that require order-$N_{L}^{2}$ operations, which may dominate the
calculation as $N_{L}$ grows. 

After each time step is completed, we update the time $t_{n}\to t_{n+1}=t_{n}+\Delta T$,
the operators $\mathcal{L}_{\alpha}\to e^{i\mathcal{H}_{0}\Delta T}\mathcal{L}_{\alpha}e^{-i\mathcal{H}_{0}\Delta T}$
($\alpha=1,\dots,N_{L}$) and $\mathcal{V}_{0}\to e^{i\mathcal{H}_{0}\Delta T}\mathcal{V}_{0}e^{-i\mathcal{H}_{0}\Delta T}$.
Using the new value of $\mathcal{L}_{\alpha}$ and $\Phi$ we calculate
$\Lambda_{\beta}$ $(\beta=0,\dots,N_{L})$ in preparation for the
next time-step. 

We set up the calculation in the following way. First we define a
macro time state $\tau=N\Delta T$. We propagate from $\Phi^{0}=\Psi\left(0\right)$
with $N$ steps of $\Delta T$, reaching $\ensuremath{\Phi^{N}}$
and then we set

\begin{align*}
\Psi^{N} & =e^{-i\mathcal{H}_{0}\tau}\Phi^{N}.
\end{align*}

It is worth mentioning that our algorithm covers as a special case
the \emph{linear unraveling procedure}\citep{appel2009stochastic,li2020exponential}, 

\begin{equation}
\left|d\psi\right\rangle =-i\left(\mathcal{H}_{0}+\mathcal{\theta}\left(t\right)\mathcal{V}_{0}\left(t\right)-i\mathcal{L}_{\alpha}^{\dagger}\mathcal{L}_{\alpha}\right)\left|\psi\right\rangle dt+\mathcal{L}_{\alpha}\left|\psi\right\rangle dw^{\alpha},\label{eq:Linear-SSE}
\end{equation}
which is obtained from Eq.~(\ref{eq:SSE}) by setting $\left\langle \mathcal{L}_{\alpha}\right\rangle \to0$.
Indeed, one can use the algorithm above and simplify by replacing
$\Lambda_{\alpha}$ by $\mathcal{L}_{\alpha}$, 'X' by $\frac{\Delta T^{2}}{2}\left|00\right\rangle +\left|dc\right\rangle $
and setting both 'Y' and 'XY' to zero in Eq.~(\ref{eq:Second-Order-Correction}). 

\begin{figure}
\begin{centering}
\includegraphics[width=0.9\columnwidth]{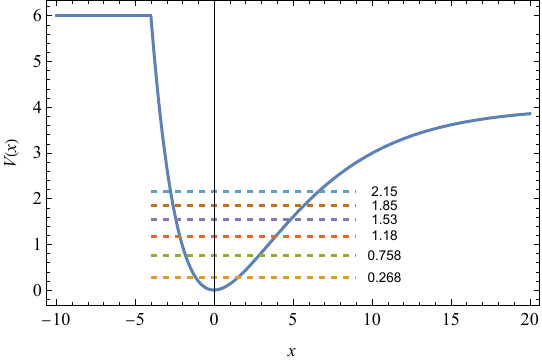}
\par\end{centering}
\caption{\label{fig:Morse-pot_and_EVs}The Morse potential $U\left(x\right)$
used in this example. The dashed lines indicate the low lying energy
eigenvalues.}
\end{figure}

\section{Validation: Morse oscillator}

The example for our method is a Morse oscillator coupled to the environment
at inverse temperature $\beta_{e}$. The particle has mass $m=1$
and the truncated Morse potential is $U\left(x\right)=\max\left[U_{max},V_{\infty}\left(1-e^{-ax}\right)^{2}\right]$
(see Fig.~\ref{fig:Morse-pot_and_EVs}), with $V_{\infty}=4$, $a=0.2$,
and $U_{max}=6$. As before, we use atomic units: $a_{0}$ (Bohr radius)
for lengths , $E_{h}$ (Hartree energy) for energy, $m_{e}$ (electron
mass) for mass and $\hbar E_{h}^{-1}$ for time. The wave functions
$\psi\left(x\right)$ we consider here may have non zero values only
in the interval $x\in\left[-10,30\right]$. We represent the system
on a 31-point grid of unit spacing ($\Delta x=1$): 
\begin{equation}
x_{n}=-10+n\Delta x,\quad n=0,\dots,30.
\end{equation}
The wave functions map into the vectors $\psi_{n}=\psi\left(x_{n}\right)$.
The position ($\mathcal{X}$) and potential $\mathcal{U}_{0}\equiv U\left(\mathcal{X}\right)$
operators operate as $\left(\mathcal{X}\psi\right)_{n}=x_{n}\psi_{n}$
and $\left(\mathcal{U}_{0}\psi\right)_{n}=U\left(x_{n}\right)\psi_{n}$
respectively. The kinetic energy operator is the finite difference
operator $\left(\mathcal{K}_{0}\psi\right)_{n}=-\frac{\hbar^{2}}{2m\Delta x^{2}}\left(\psi_{n-1}-2\psi_{n}+\psi_{n+1}\right)$,
combined with the boundary condition $\psi_{-1}\equiv\psi_{31}\equiv0$.
This defines the Hamiltonian $\mathcal{H}_{0}=\mathcal{K}_{0}+\mathcal{U}_{0}$.
The lowest lying bound energy levels of this Hamiltonian, determined
by diagonalization, are given in Fig.~\ref{fig:Morse-pot_and_EVs}.

We take only two Lindblad operators 
\begin{equation}
\mathcal{L}_{\pm\omega_{B},\mathcal{T}}=\sqrt{\gamma_{\pm\omega_{B}}}\times\frac{1}{2\mathcal{T}}\int_{-\mathcal{T}}^{\mathcal{T}}e^{\pm i\omega_{B}\tau}\mathcal{X}_{H}\left(\tau\right)d\tau\label{eq:Morse-lindblad-matrix}
\end{equation}
where $\mathcal{X}_{H}\left(\tau\right)=e^{\frac{i}{\hbar}\mathcal{H}_{0}\tau}\mathcal{X}e^{-\frac{i}{\hbar}\mathcal{H}_{0}\tau}$
is the time-dependent Heisenberg operator for $\mathcal{X}$, $\mathcal{T}=10$
and $\omega_{B}=\hbar^{-1}\left(E_{1}-E_{0}\right)=\hbar^{-1}\times0.4903$.
The rates in Eq.~(\ref{eq:Morse-lindblad-matrix}) are chosen as
\[
\gamma_{\pm\omega_{B}}=\frac{\gamma_{0}}{1+e^{\pm\beta_{e}\hbar\omega_{B}}}
\]
where $\gamma_{0}=0.2$ and the environment inverse temperature $\beta_{e}=4$.
These rates obey the detailed balance condition 
\begin{equation}
\frac{\gamma_{\omega_{B}}}{\gamma_{-\omega_{B}}}=e^{-\beta_{e}\hbar\omega_{B}}.\label{eq:detailed-balance}
\end{equation}

\begin{figure*}
\begin{centering}
\includegraphics[width=0.52\textwidth]{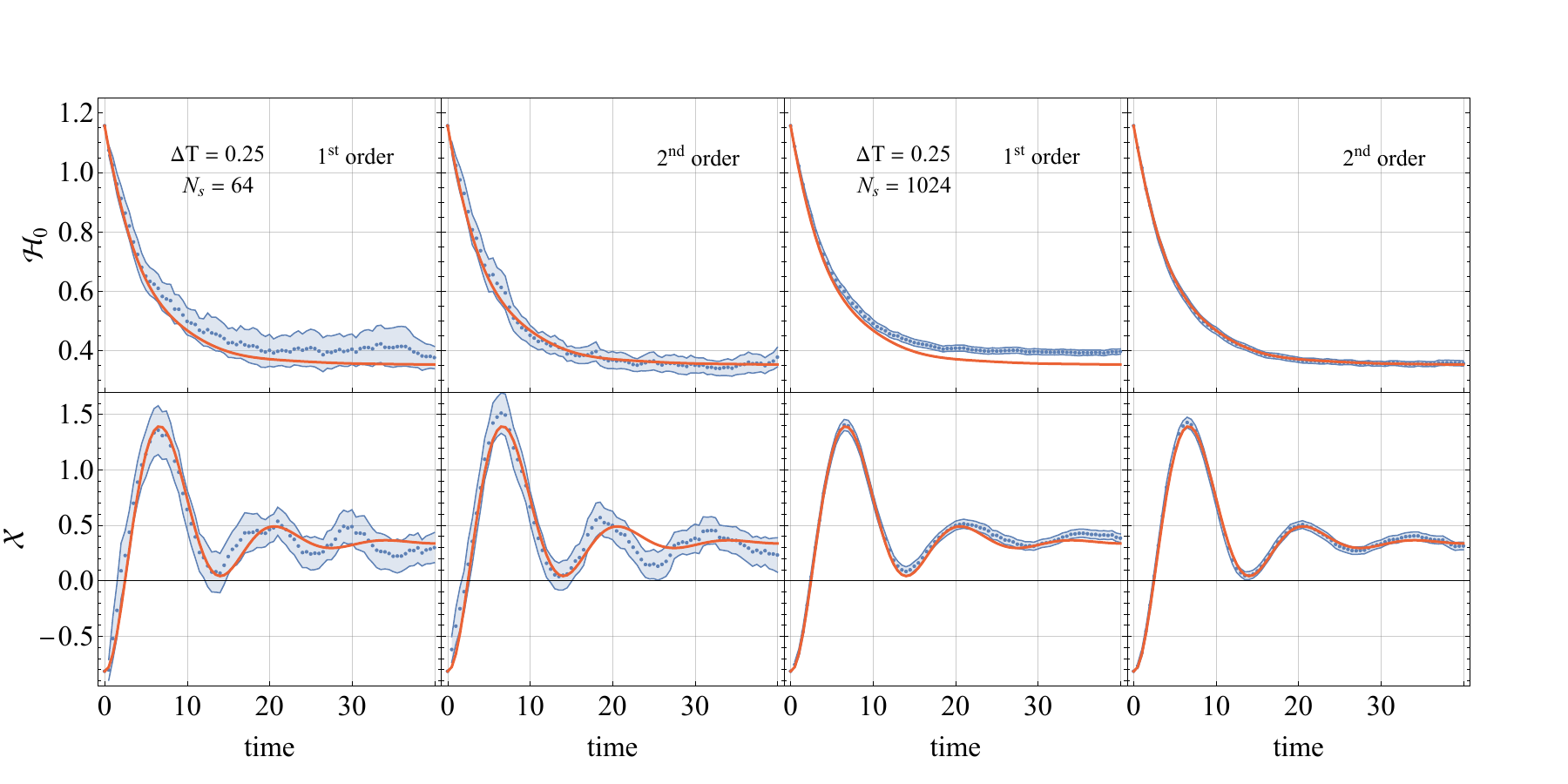}\includegraphics[width=0.52\textwidth]{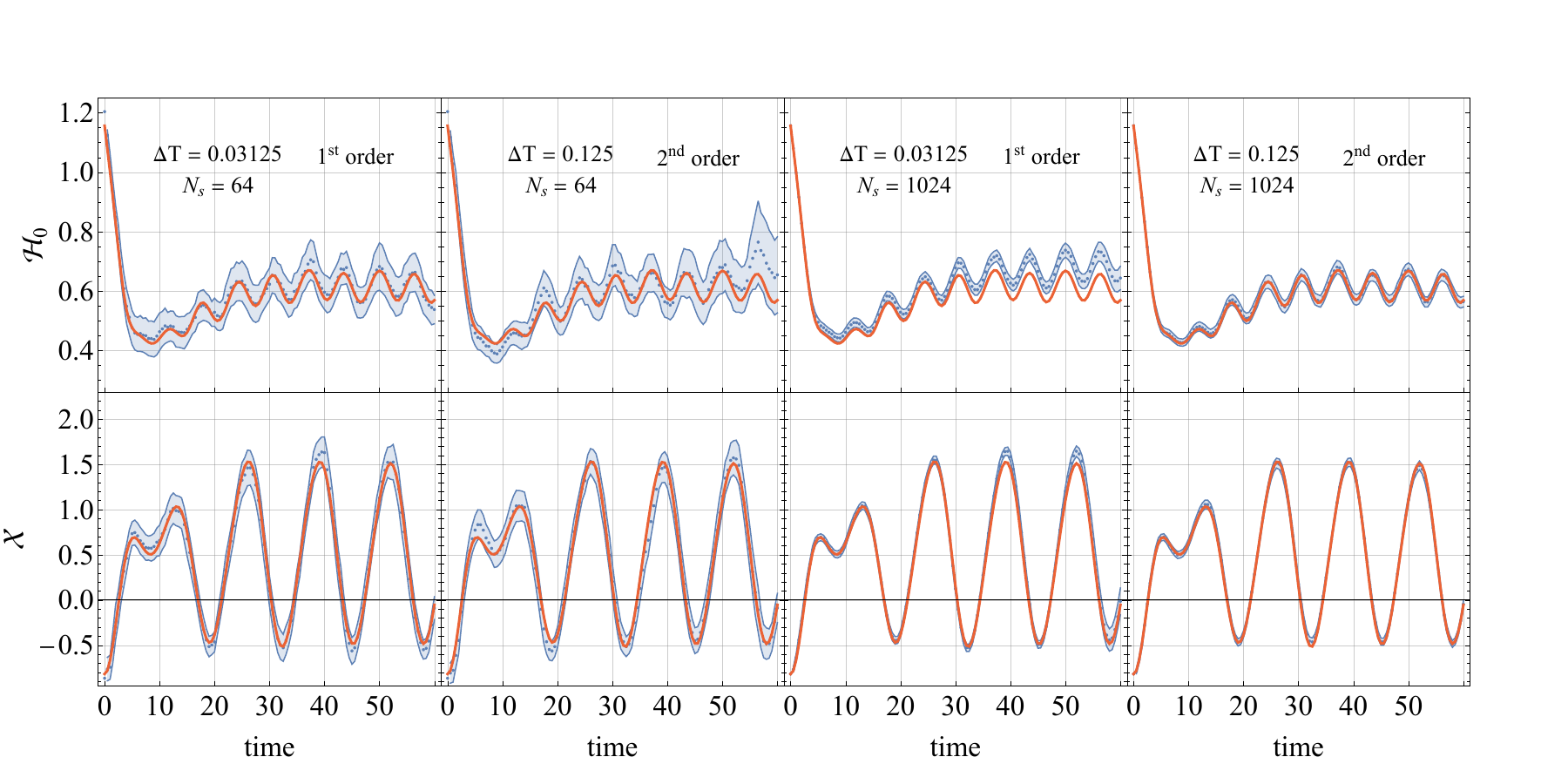}
\par\end{centering}
\caption{\label{fig:CoolingTransients}The 95\% confidence region (blue shade)
for the energy (top panels) and position (bottom panels) transients
of the free (left) and driven (right) Morse oscillator starting from
a hot state, obtained from the first- and second-order solutions of
the ISE (Eqs.~(\ref{eq:First-Order-Correction}) and (\ref{eq:Second-Order-Correction}))
using $N_{s}=64$ and $1024$ samples. Also shown, as red lines, the
numerically exact energy and position transients calculated by solving
Eq.~(\ref{eq:Lindbald}). }
\end{figure*}

The last element of the model problem is the initial state, which
we take as a pure state $\rho\left(0\right)=\left|\xi\right\rangle \left\langle \xi\right|$:
\begin{align}
\left|\xi\right\rangle  & =\frac{1}{\sqrt{3}}\left(\left|\psi_{2}\right\rangle +\left|\psi_{3}\right\rangle +\left|\psi_{4}\right\rangle \right)\label{eq:initial-state}
\end{align}
 where $\left|\psi_{n}\right\rangle $ are the eigenvectors of the
Hamiltonian operator $\mathcal{H}_{0}$. 

\subsection{The 'free' oscillator}

We first discuss a time-independent case, where the oscillator is
free, i.e., is not subjected to an external driving force beyond the
interaction with the environment. Using a small time step and a fourth-order
Runge-Kutta propagator, we evolve the density operator according to
the Lindblad Equation (Eq.~\ref{eq:Lindbald}), starting from $\rho\left(t=0\right)$
and obtain highly accurate reference values for benchmarking the stochastic
propagators. We find that the extended time limit of the evolved state
is close, but not exactly equal, to the thermal state at the environment
temperature. In order to converge fully into the thermal state, we
need to provide more Lindblad operators than just the two we consider
here.  

\begin{figure}
\centering{}\includegraphics[width=1\columnwidth]{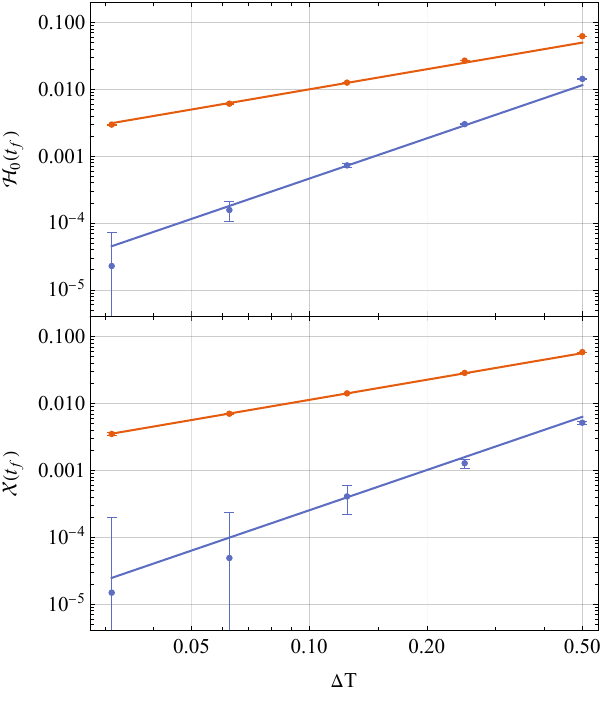}\caption{\label{fig:cooling-error}Log-plots showing the first- (red) and second-
(blue) order confidence intervals for the energy $\mathcal{H}_{0}\left(t_{f}\right)$
(top panel) and position $\mathcal{X}\left(t_{f}\right)$ (bottom
panel) expectation values at $t_{f}=7$ vs the time step $\Delta T$.
Dotted straight lines show asymptotic first- and second- order behavior
(Eq.~(\ref{eq:weak-convergence})). The number of samples used for
estimating the confidence intervals was $N_{s}=64\times10^{6}$.}
\end{figure}

The stochastic calculation provides confidence intervals for the Lindblad
expectation values $\text{Tr}\left[\rho_{t}\mathcal{A}\right]$ of
any given observable of interest $\mathcal{A}$. The procedure is
a straightforward application of statistical analysis. We run our
propagator $N_{s}$ times (with independent random numbers) collecting
$N_{s}$ samples of quantum expectation values $A_{t}^{\left(k\right)}\equiv\left\langle \Psi_{t}^{\left(k\right)}\left|\mathcal{A}\right|\Psi_{t}^{\left(k\right)}\right\rangle /\left\langle \Psi_{t}^{\left(k\right)}\left|\Psi_{t}^{\left(k\right)}\right.\right>$
($k=1,\dots,N_{s}$) and then construct the 95\% confidence interval
as $\left[\bar{A}_{t}-\Delta A_{t},\bar{A}_{t}+\Delta A_{t}\right]$,
where $\bar{A}_{t}$ is the sample average, $\Delta A_{t}=2\times S_{t}/\sqrt{N_{s}}$
is the interval width, and $S_{t}$ is the sample standard deviation.
The factor 2 is the large sample t-factor corresponding to a confidence
level of $\sim95\%$. In Fig.~\ref{fig:CoolingTransients}, we show
confidence intervals for two observables, the energy $\mathcal{H}_{0}$
and the position $\mathcal{X}$, using $N_{s}=64$ and $1024$ samples
based on the first- and second-order propagators with time step $\Delta T=0.25$.
For reference, the figure also shows, as a red solid line, the numerically
exact expected value $\mathbb{E}\left[\left\langle \psi_{t}\left|\mathcal{A}\right|\psi_{t}\right\rangle \right]=\text{Tr}\left[\rho_{t}\mathcal{A}\right]$. 

The first-order calculation exhibits a noticeable energy bias at $\Delta T=0.25$,
even when the confidence interval is broad (when $N_{s}=64$). At
the same time, the bias from the second-order calculation is not noticeable
even for $N_{s}=1024$ sampling. We discuss the weak order convergence
comparing first- and second-order methods below. The standard deviation
$S_{t}$ in the first- and second-order calculations is around 0.08
for the energy and 0.6 for the position; interestingly, it does not
grow with time. We also checked the algorithm for the case of over-damped
dynamics where a parameter of $\gamma=0.6$ was used. We observed
similar trends as in weak coupling in terms of the accuracy of the
calculation (see details in the supplementary material). In terms
of stability, both first- and second- order calculations required
time steps of at least 0.0625, for larger time steps the solution
was unstable and diverged.

In a weak order-$o$ method $\mathbb{E}\left[\left\langle \Psi_{t}\left|\mathcal{A}\right|\Psi_{t}\right\rangle \right]$
should approach the exact value $\text{Tr}\left[\rho_{t}\mathcal{A}\right]$
as the $o$ power of $\Delta T$. More precisely, there exist $\Delta T_{0}>0$
and $C>0$ such that: 
\begin{equation}
\Delta T<\Delta T_{0}\Rightarrow\left|\mathbb{E}\left[\left\langle \Psi_{t}\left|\mathcal{A}\right|\Psi_{t}\right\rangle \right]-\text{Tr}\left[\rho_{t}\mathcal{A}\right]\right|\le C\times\Delta T^{o}.\label{eq:weak-convergence}
\end{equation}
To test whether this condition is obeyed we need to know $\mathbb{E}\left[\left\langle \Psi_{t}\left|\mathcal{A}\right|\Psi_{t}\right\rangle \right]$,
and this not available directly. However we can build a very small
95\% confidence interval by extensive sampling (taking $N_{s}=64\times10^{6}$),
as shown in Fig.~\ref{fig:cooling-error} for the energy and position
observables at time $t_{f}=7$ as function of the time step $\Delta T$.
The asymptotic behavior of Eq.~(\ref{eq:weak-convergence}) is clearly
seen as the asymptotic lines do indeed fit through the very small
confidence intervals. The power of the second-order calculations is
also evident as its error with $\Delta T=0.25$ is smaller than the
error in the first-order calculation using a time step smaller by
a factor 8. 

To assess the utility of the second-order vs the first-order solvers,
we note that for the example given here, the wall-time for the former
is only a 1.5 times larger than the latter. This small ratio in wall-times
will characterize larger systems, as long as there is only one Lindblad
operators. From the discussion above, concerning the time-step (and
hence number of time steps) required by both methods we conclude that
in the present example, the second-order solver is five times more
efficient than the first-order one, for low-accuracy calculations.
For higher accuracies, it is considerably more efficient. However,
the wall time in the second-order calculation depends quadratically
on the number $N_{L}$ of Lindblad operators, while that of the first-order
is linear in $N_{L}$. Hence, the numerical cost of the second-order
calculation may exceed that of the first-order calculation as $N_{L}$
grows. 

We mention briefly that linear unraveling (Eq.~(\ref{eq:Linear-SSE}))
has a variance one to two orders of magnitude larger than for the
nonlinear unraveling (and it grows linearly with time). Hence, the
nonlinear unraveling is expected to be superior in actual applications. 

\subsection{The driven oscillator}

In this example, we subject the Morse oscillator to a driving time-dependent
field
\begin{equation}
\mathcal{V}_{0}\theta\left(t\right)=\mathcal{X}\mathcal{F}\sin\left(\omega t\right)
\end{equation}
with $\mathcal{F}=0.2$ and $\omega=0.49$. The frequency is resonant
between the ground and the first excited states of the oscillator.
In Fig.~\ref{fig:CoolingTransients} (right) we show first- and second-order
results for $N_{s}=64$ and $1024$ samples. The oscillator starts
from the same pure state as in the example of the previous section
(see Eq.~(\ref{eq:initial-state})). Under the driving force it strives
to cool due to the interaction with the cold environment but the driving
field acts to heat it. Eventually, a quasi-stationary non-thermal
state forms, with the oscillator energy and position oscillating strongly
in time. The first-order solution is unstable for $\Delta T>0.03125$
and even at this small time-step exhibits a large energy bias (red
line not passing in the confidence interval for $N_{s}=1024$). The
second-order results are stable and much more accurate even when $\Delta T=0.125$.
As for the standard deviation $S_{t}$ in the driven oscillator, it
is around 0.25 for energy and 0.6 for position. As with the free oscillator,
$S_{t}$ does not grow with time. 

\section{Conclusions}

We have presented a weak second-order method for solving the It\^{o}-Schr\"{o}dinger
equation related to quantum state diffusion unraveling of the Lindblad
equation. One of the approach's critical characteristics is working
in the interaction picture, helping stability and accuracy even for
relatively large time steps. Another significant characteristic of
our approach is nonlinear unraveling, using within the equation the
expectation value of the Lindblad operator, which reduces the variance
(in comparison to the linear unraveling schemes). Moreover, the use
of \emph{explicitly normalized }expectation values of the Lindblad
operators (Eq.~(\ref{eq:ExpectationValue})), further stabilizes
the propagation. Another characteristic of our approach is using exact
derivatives, which are readily available since our nonlinearity is
analytical, for the It\^{o}-Taylor expansion (as opposed to other second-order
approaches, such as the Runge-Kutta method, which bypasses derivatives
using finite difference). Lastly, our method uses complex Wiener processes.

We have tested the method on the problem of cooling an initially hot
Morse oscillator coupled to a colder environment. We studied both
a free and a driven oscillator. In both cases, we showed good accuracy
of the second-order method when the time step was $\Delta T\omega_{B}\approx0.1$
or smaller, achieving useful confidence intervals with a relatively
small amount of sampling. 

We have used 1D examples to benchmark our methods. For such small
systems, unraveling does not save computational resources relative
to a complete solution of the Lindblad equation. However, the latter
method has cubic scaling in wall time and quadratic scaling in memory,
and therefore, unraveling can become more efficient as systems grow.
One clear advantage of unraveling is that it does not require storing
the density matrix, saving a vast amount of computer memory. Furthermore,
the most intensive part of the unraveling calculation, namely transforming
to and from the interaction picture, can be accomplished by iterative
methods \citep{feit1982solution,kosloff1988timedependent} involving
a fixed number of Hamiltonian applications to any given ket. As systems
grow, this latter operation becomes linear-scaling in complexity,
endowing the entire unraveling procedure with the same complexity.
Thus, there is a massive reduction in computational time relative
to a complete solution of the Lindblad equation in the limit of large
systems. Furthermore, multiprocessor parallelization can easily overcome
the burden of repeated sampling in the unraveling procedure.

The propagator developed in the present paper is our first step towards
a more general goal of constructing a framework for studying quantum
decoherence and dissipation in large molecular and nanoscale systems.
The computational wall-time involved in the second-order calculation
scales quadratically with the number $N_{L}$ of Lindblad operators.
Therefore, our immediate future work will involve a method to contract
Lindblad operators so that a small, hopefully, system-size-independent
number of operators can be used. In addition, in the future, we may
try to develop solvers for stochastic Schr\"{o}dinger equations that unravel
non-Markovian master equations. Such solvers are required since the
Markovian dynamics may result in unreliable predictions of bath-induced
coherences \citep{leijnse2008kinetic,esposito2015efficiency,gao2016simulation,kidon2015exactcalculation}).

\paragraph{Supplementary Material}

Supplementary material is given on the derivation of Eq.~(\ref{eq:Second-Order-Correction})
and on the results of the Morse oscilator in the overdamped limit.

\paragraph{Acknowledgments}

The authors gratefully acknowledge funding from the Israel Science
Foundation grant number ISF-800/19.

\bibliographystyle{unsrt}

\end{document}